\documentclass[
    ,final            
  ]
  {aipproc}

\layoutstyle{8x11double}

\def\mn={MNRAS}

\begin{document}

\title{Two populations are better than one: Short gamma-ray bursts from SGR giant flares \textit{and} NS-NS mergers}

\classification{98.70.Rz, 97.60.Jd}
\keywords      {gamma-ray bursts, magnetars, SGR, giant flares, neutron star binaries}

\author{Robert Chapman}{
  address={Centre for Astrophysics Research, University of Hertfordshire, College Lane, Hatfield AL10~9AB, UK}
}

\author{Robert S. Priddey}{
  address={Centre for Astrophysics Research, University of Hertfordshire, College Lane, Hatfield AL10~9AB, UK}
}

\author{Nial R. Tanvir}{
  address={Department of Physics and Astronomy, University of Leicester, Leicester, LE1~7RH, UK}
}

\begin{abstract}
With a peak luminosity of $\sim10^{47}~\rm{erg~s^{-1}}$, the December 27th 2004 giant flare from SGR1806-20 would have been visible by BATSE (the Burst and Transient Source Experiment) out to $\approx50~\rm{Mpc}$\citep{2007PhR...442..166N,2006MPLA...21.2171T}. It is thus plausible that some fraction of the short duration Gamma-Ray Bursts (sGRBs) in the BATSE catalogue were due to extragalactic magnetar giant flares. According to the most widely accepted current models, the remaining BATSE sGRBs were most likely produced by compact object (neutron star-neutron star or neutron star-black hole) mergers with intrinsically higher luminosities \citep{2007PhR...442..166N}. Previously, by examining correlations on the sky between BATSE sGRBs and galaxies within 155~Mpc, we placed limits on the proportion of nearby sGRBs \citep{2005Natur.438..991T}. Here, we examine the redshift distribution of sGRBs produced by assuming both one and two populations of progenitor with separate Luminosity Functions (LFs). Using the local Galactic SGR giant flare rate and theoretical NS-NS merger rates evolved according to well-known Star Formation Rate parameterisations, we constrain the predicted distributions by BATSE sGRB overall number counts. We show that only a dual population consisting of both SGR giant flares and NS-NS mergers can reproduce the likely local distribution of sGRBs as well as the overall number counts. In addition, the best fit LF parameters of both sub-populations are in good agreement with observed luminosities. 
\end{abstract}


\maketitle


\section{Introduction}
The leading candidate progenitor model for short duration Gamma-Ray Bursts (sGRBs) is the merger of two compact objects, neutron star-neutron star (NS-NS) or neutron star-black hole~\citep{2007PhR...442..166N}. However the initial spike in a giant flare from a Soft Gamma Repeater (SGR) in a relatively nearby galaxy would also appear as a sGRB. For example, the December 27th 2004 event from SGR1806-20 would have been visible by BATSE out to $\approx50~\rm{Mpc}$~\citep{2007PhR...442..166N,2006MPLA...21.2171T}. Thus it is plausible that some fraction of sGRBs are extragalactic SGR giant flares. Previously,~\citet{2005Natur.438..991T} demonstrated that between 10 and 25 per cent of BATSE sGRBs were correlated on the sky with galaxies within $\approx113~\rm{Mpc}$. The Luminosity Function (LF) of sGRBs has been investigated previously assuming a single population~\citep{Guetta:2004fc} in order to determine the intrinsic sGRB rate and most likely LF parameters. Here we assume intrinsic rates given by both the observed Galactic SGR flare rates and modelled NS-NS merger rates in order to investigate single and dual population LF parameters. Obviously there are significant uncertainties in these rates: the Galactic giant flare rate in particular is estimated from only 3 observed events. Regardless of these uncertainties and the exact form of LF for a second population, we find a single progenitor population constrained by overall number counts cannot produce sufficient local events. We present preliminary results that show only a dual population can reproduce the likely local sGRB distribution as well as the overall number counts.

\section{Methods}
The number of sGRBs, $N$, observed above threshold $p$ in time $T$ and solid angle $\Omega$ is given by equation~\ref{lfeq}, where $\Phi(L)$ is the sGRB LF, $R_{GRB}(z)$ is the sGRB event rate at redshift $z$, $dV(z)/dz$ is the comoving volume element at $z$ and $z_{max}$ for a burst of luminosity $L$ is determined by the detector flux threshold and the luminosity distance of the event. We assume a spectral photon index of $–1.1$ in the BATSE energy range to convert photon flux to energy flux.

\begin{equation}
N(>p)=\frac{\Omega T}{4\pi}\int_{L_{min}}^{L_{max}}\Phi(L)dl\int_{0}^{z_{max}}\frac{R_{GRB}(z)}{1+z}\frac{dV(z)}{dz}dz \label{lfeq}
\end{equation}

The sGRB rate, $R_{GRB(z)}$ is given by equation~\ref{rgrb} where $\rho_{GRB}$ is the rate of sGRBs per progenitor, $\rho_{P}(0)$ is the intrinsic progenitor formation rate and $F(z)$ describes the progenitor production rate as a function of $z$. 

\begin{equation}
R_{GRB}=\rho_{GRB} \times \rho_{P}(0) \times F(z) \label{rgrb}
\end{equation}

For NS-NS mergers, $\rho_{GRB}=1$ and $F(z)$ is the delayed rate given by the convolution of the Star Formation Rate at $z$ (SFR(z)) with a distribution of delay times to represent merger formation. For SGR flares, $\rho_{GRB}=3\times10^{-2}$ (the observed local rate of giant flares per Galactic SGR) and $F(z)$ follows both SFR(z) for magnetar production from supernovae and the delayed SFR(z) to allow for production by White Dwarf binary mergers \citep{2006MNRAS.368L...1L}. We take SFR(z) as given by the SF2 model of~\citet{2001ApJ...548..522P}, with the local SFR from~\citep{1995ApJ...455L...1G}. $\rho_{P(0)}$ SGR rates per Milky Way equivalent galaxy are from~\citet{2006MNRAS.368L...1L}, and we follow~\citet{2007PhR...442...75K} for a NS-NS merger rate of $10^{-5}~\rm{yr}^{-1}$.


LFs for SGR giant flares and NS-NS mergers are not well constrained. A log-normal LF approximates the theoretical NS-NS merger distribution~\citep{2003MNRAS.343L..36R}, and is also plausible for SGR giant flares. Here we assume sGRBs (of differing luminosities) to arise from both SGRs and NS-NS mergers. We also investigate a single merger population with Schechter function and log-normal LFs. Merger time distributions are similarly unknown, and we assume $P(\rm{log}(\tau))d\rm{log}(\tau)=constant$ between $10^7$ and $10^{10}$ years, approximating theoretical distributions~\citep{2006ApJ...648.1110B}.

\begin{figure}[!htbp]
  \includegraphics[height=.2\textheight]{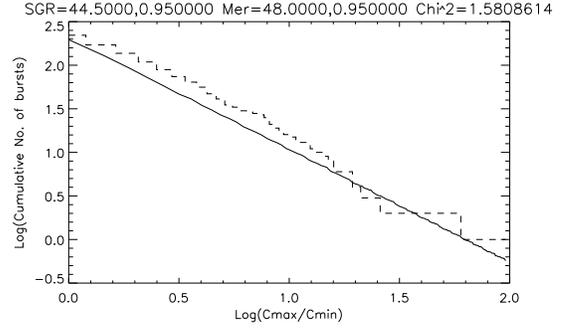}
  \caption{Example of best fit plot of number counts predicted from Equation~\ref{lfeq} using dual population lognormal LFs versus observed $C_{max}/C_{min}$. Fits are performed for differential number counts, but shown cumulative in the figure. Dashed line is observed $C_{max}/C_{min}$, solid line predicted number counts. Reduced $\chi_2=1.58$.\label{bestfit}}
\end{figure}
\begin{figure}[!htbp]
  \includegraphics[height=.2\textheight]{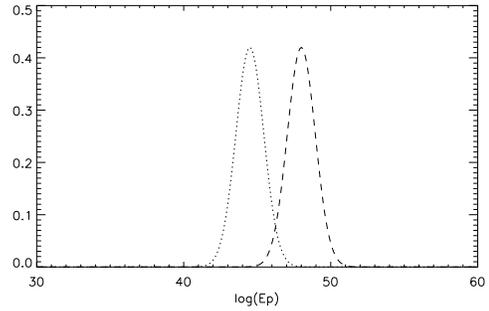}
  \caption{Dual population best fit log-normal Luminosity Functions from Figure~\ref{bestfit}. Dotted line is LF for bursts due to SGR giant flares, median $log(E_p)=44.5, \sigma_{log}=0.95$. Dashed line is LF for bursts due to NS-NS mergers $log(E_p)=48.0, \sigma_{log}=0.95$. Other details as Figure~\ref{bestfit}.\label{lf}}
\end{figure}
\begin{figure}[!htbp]
  \includegraphics[height=.2\textheight]{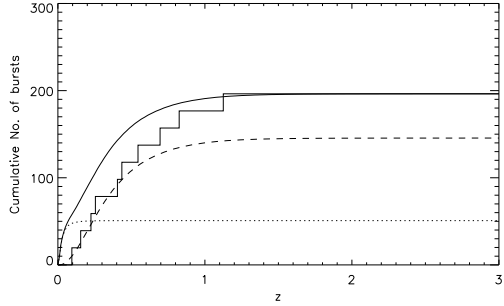}
  \caption{Cumulative distribution of bursts out to $z=3$ using best fit dual population log-normal LFs results from Figure~\ref{lf}. Also shown (solid line) is the cumulative distribution of the handful of sGRBs with known redshifts~\citep{2007astro.ph..2694B}. Other details as Figure~\ref{bestfit}.\label{lognormallz}}
\end{figure}

The $C_{max}/C_{min}$ table from the BATSE catalogue\citep{1999ApJS..122..465P} provides peak count rate for bursts in units of threshold rate. The BATSE threshold was varied historically, and to analyse a consistent set of bursts we restricted the table to only those sGRBs recorded when the threshold was set to $5.5\sigma$ above background in at least 2 detectors in the $50-300~\rm{keV}$ range. The all sky equivalent period this represents is $\sim1.3~\rm{years}$. By varying the parameters of the LFs (median and sigma for log-normal, slope and $L_{*}$ for Schechter) and fitting the predicted $N(>p)$ to the $C_{max}/C_{min}$ distribution, we found the best fit LF parameters by $\chi^{2}$ minimization. Figure~\ref{bestfit} shows an example of such a fit. Having obtained the best fit LF parameters, the redshift distribution of predicted sGRBs can be calculated, as shown in Figures~\ref{lf}-\ref{onlyschecter150Mpc}.

\section{Results and Conclusions}

\begin{figure}[!htbp]
  \includegraphics[height=.2\textheight]{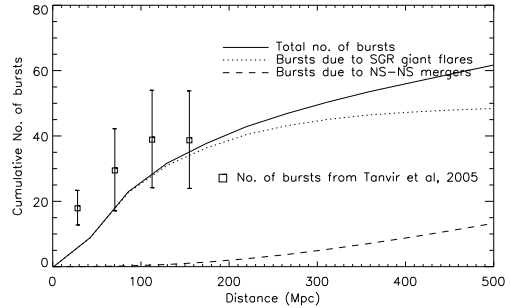}
  \caption{Cumulative distribution of bursts within 500Mpc using best fit dual population log-normal LFs results from Figure~\ref{lf}. Also shown (open squares with $1\sigma$ errors) are the rates of local sGRBs as measured by correlation analyses with local galaxies from~\citet{2005Natur.438..991T}. Other details as Figure~\ref{bestfit}.\label{lognorm150Mpc}}
\end{figure}

\begin{figure}[!htbp]
  \includegraphics[height=.2\textheight]{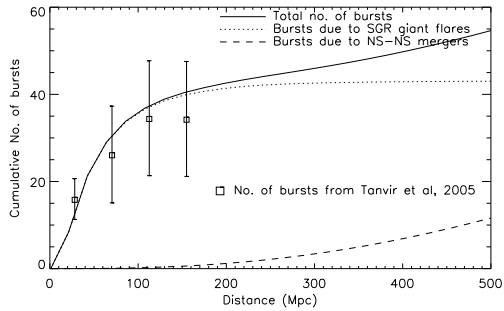}
  \caption{Cumulative distribution of bursts within 500Mpc using best fit dual population with log-normal LF for the SGR flares (median $log(E_p)=45.0, \sigma_{log}=0.75$), and a Schechter function ($L_{*}=50.0, slope=-0.9$) for the NS-NS merger population. Reduced $\chi_2=2.36$, other details as Figure~\ref{bestfit}.\label{schecter150Mpc}}
\end{figure}

\begin{figure}[!htbp]
  \includegraphics[height=.2\textheight]{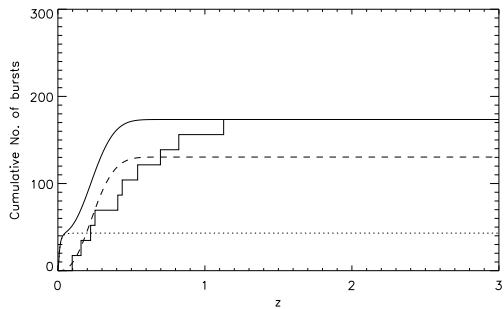}
  \caption{Cumulative distribution of bursts out to $z=3$ using best fit dual population (lognormal SGR, Schechter mergers) as Figure~\ref{schecter150Mpc}. Other details as Figure~\ref{lognormallz}.\label{schecterallz}}
\end{figure}

\begin{figure}[!htbp]
  \includegraphics[height=.2\textheight]{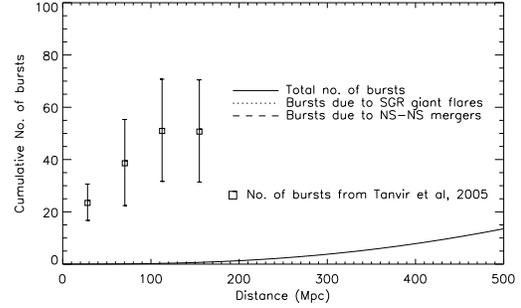}
\caption{Cumulative distribution of bursts within 500Mpc using best fit single Schechter function ($L_{*}=50.25, slope=-0.9$, reduced $\chi_2=2.53$.) for the NS-NS merger population. Other details as Figure~\ref{lognorm150Mpc}.\label{onlyschecter150Mpc}}
\end{figure}

Although a single NS-NS merger population luminosity function can produce reasonable fits to the overall number counts, only samples with two populations of differing luminosity distributions reproduce the likely local population of sGRBs (Figures~\ref{lognorm150Mpc} and \ref{schecter150Mpc}) - even Schechter type single luminosity functions, which are dominated by low-luminosity events, fail to reproduce the nearby population (Figure~\ref{onlyschecter150Mpc}). Furthermore, the overall redshift distribution of the dual population samples are consistent with the handful of redshifts known for sGRBs\citep{2007astro.ph..2694B}. Despite large uncertainties in assumed intrinsic rates, the best fit LF parameters also agree well with the few peak luminosities currently known for both SGR giant flares and cosmological sGRBs.   

\begin{theacknowledgments}
RC and RSP are grateful for the support of the University of Hertfordshire. NRT acknowledges the support of a UK STFC senior research fellowship.
\end{theacknowledgments}


\bibliographystyle{aipproc}   


\bibliography{bobrefs}

\begin{thebibliography}{12}
\expandafter\ifx\csname natexlab\endcsname\relax\def\natexlab#1{#1}\fi
\providecommand{\enquote}[1]{``#1''}
\expandafter\ifx\csname url\endcsname\relax
  \def\url#1{\texttt{#1}}\fi
\expandafter\ifx\csname urlprefix\endcsname\relax\def\urlprefix{URL }\fi
\providecommand{\eprint}[2][]{\url{#2}}

\bibitem[{Nakar}(2007)]{2007PhR...442..166N}
E.~{Nakar}, \emph{\physrep} \textbf{442}, 166--236 (2007),
  \eprint{arXiv:astro-ph/0701748}.

\bibitem[{Taylor} and {Granot}(2006)]{2006MPLA...21.2171T}
G.~B. {Taylor}, and J.~{Granot}, \emph{Modern Physics Letters A} \textbf{21},
  2171--2188 (2006), \eprint{arXiv:astro-ph/0609595}.

\bibitem[{Tanvir} et~al.(2005)]{2005Natur.438..991T}
N.~R. {Tanvir}, R.~{Chapman}, A.~J. {Levan}, and R.~S. {Priddey}, \emph{\nat}
  \textbf{438}, 991--993 (2005), \eprint{arXiv:astro-ph/0509167}.

\bibitem[Guetta and Piran(2005)]{Guetta:2004fc}
D.~Guetta, and T.~Piran, \emph{Astrophys. J.} \textbf{619}, 412--419 (2005),
  \eprint{astro-ph/0407429}.

\bibitem[{Levan} et~al.(2006)]{2006MNRAS.368L...1L}
A.~J. {Levan}, G.~A. {Wynn}, R.~{Chapman}, M.~B. {Davies}, A.~R. {King}, R.~S.
  {Priddey}, and N.~R. {Tanvir}, \emph{\mnras} \textbf{368}, L1--L5 (2006),
  \eprint{arXiv:astro-ph/0601332}.

\bibitem[{Porciani} and {Madau}(2001)]{2001ApJ...548..522P}
C.~{Porciani}, and P.~{Madau}, \emph{\apj} \textbf{548}, 522--531 (2001),
  \eprint{arXiv:astro-ph/0008294}.

\bibitem[{Gallego} et~al.(1995)]{1995ApJ...455L...1G}
J.~{Gallego}, J.~{Zamorano}, A.~{Aragon-Salamanca}, and M.~{Rego}, \emph{\apjl}
  \textbf{455}, L1+ (1995).

\bibitem[{Kalogera} et~al.(2007)]{2007PhR...442...75K}
V.~{Kalogera}, K.~{Belczynski}, C.~{Kim}, R.~{O'Shaughnessy}, and B.~{Willems},
  \emph{\physrep} \textbf{442}, 75--108 (2007),
  \eprint{arXiv:astro-ph/0612144}.

\bibitem[{Rosswog} and {Ramirez-Ruiz}(2003)]{2003MNRAS.343L..36R}
S.~{Rosswog}, and E.~{Ramirez-Ruiz}, \emph{\mnras} \textbf{343}, L36--L40
  (2003).

\bibitem[{Belczynski} et~al.(2006)]{2006ApJ...648.1110B}
K.~{Belczynski}, R.~{Perna}, T.~{Bulik}, V.~{Kalogera}, N.~{Ivanova}, and D.~Q.
  {Lamb}, \emph{\apj} \textbf{648}, 1110--1116 (2006),
  \eprint{arXiv:astro-ph/0601458}.

\bibitem[{Paciesas} et~al.(1999)]{1999ApJS..122..465P}
W.~S. {Paciesas}, C.~A. {Meegan}, G.~N. {Pendleton}, M.~S. {Briggs},
  C.~{Kouveliotou}, T.~M. {Koshut}, J.~P. {Lestrade}, M.~L. {McCollough}, J.~J.
  {Brainerd}, J.~{Hakkila}, W.~{Henze}, R.~D. {Preece}, V.~{Connaughton}, R.~M.
  {Kippen}, R.~S. {Mallozzi}, G.~J. {Fishman}, G.~A. {Richardson}, and
  M.~{Sahi}, \emph{\apjs} \textbf{122}, 465--495 (1999).

\bibitem[{Berger}(2007)]{2007astro.ph..2694B}
E.~{Berger}, \emph{ArXiv Astrophysics e-prints}  (2007),
  \eprint{astro-ph/0702694}.

\end{thebibliography}


\end{document}